# Measuring the Knowledge Base

# A Program of Innovation Studies

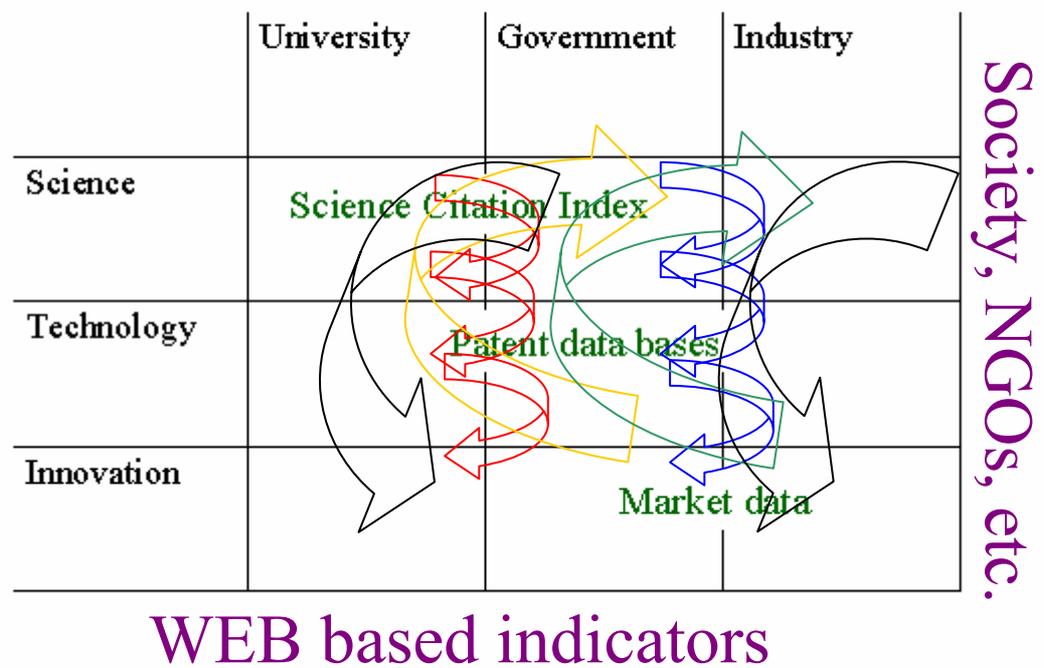

*Loet Leydesdorff*
*Andrea Scharnhorst*

MEASURING THE KNOWLEDGE BASE:
A PROGRAM OF INNOVATION STUDIES

@ Loet Leydesdorff [1] & Andrea Scharnhorst [2], 2003

1. Amsterdam School of Communications Research (ASCoR), University of Amsterdam, http://ww.leydesdorff.net
2. Networked Research and Digital Information, The Netherlands Institute for Scientific Information Services, Royal Academy of Arts and Sciences (KNAW), Amsterdam, http://ww.niwi.knaw.nl/nerdi

This report was written for the "Förderinitiative *Science Policy Studies*" of the German Bundesministerium für Bildung und Forschung. The program was organized by Rainer Hohlfeld of the Berlin-Brandenburgische Akademie der Wissenschaften. The authors acknowledge the support of these organizations.

Amsterdam, March 2003.

# Table of Contents

# SUMMARY

In a reflection of new developments in science and society, Gibbons *et al.* (1994) proposed to make a distinction between "Mode 2" and "Mode 1" types of knowledge production. Whereas "Mode 1" refers to the traditional shape of science, largely confined within institutional settings, "Mode 2" is communication driven. Organized knowledge production can then be considered as the codification of communication. Communications leave traces that can be studied as indicators. Institutions can be considered as retention mechanisms functional for the reproduction of ever more complex, that is, scientific and knowledge-based, communications.

From this perspective, "national systems of innovation" compete as niches in terms of their problem-solving capacities (Nelson & Winter, 1982). In the Triple Helix model the institutions are analyzed in terms of university-industry-government relations. The network is continuously reshaped by knowledge-based innovations that result from inventions at one level and feedback at other network levels. Knowledge-based communications and their re-combinations thus drive the institutional reform of the political economy into a knowledge-based economy. The networks differentiate further while importing knowledge in the form of innovations. As innovations take place at interfaces, the competitive advantages in a knowledge-based economy can no longer be attributed to a single node in the network.

The networks coordinate the sub-dynamics of (i) wealth production, (ii) organized novelty production, and (iii) private appropriation versus public control. Boundary-spanning mechanisms can be expected to change systems by providing new options for innovation. The Internet can be considered as a boundary-spanning mechanism at the global level. It relates different parts of the knowledge production, diffusion, and control system to one another. Academic, economic, and public spheres can use the same media for the representations. The *expectation* of global exchange relations ("globalization") changes the knowledgeable options within the lower-level systems by making new "variation" and other "selection" criteria possible.

The focus on communication enables us to operationalize the research questions in terms of indicators by using the mathematical theory of communication. For example, the systems of innovation can be measured in terms of interfaces among communications about new products and/or technological processes. Are the innovations under study incidental or systemic? The degree of systemic behaviour can be tested correlating different types of data. For example, systemness can be compared with the historical development of time-series data. Are the emerging densities in network relations also reproduced?

Two theories of communication provide the heuristics: (i) Luhmann's (1984 and 1990) sociological theory of communication with its emphasis on functional differentiation provides hypotheses, and (ii) Shannon's (1948) mathematical theory of communication can be used for the operationalization. The combination of these two theories with a very different status—i.e., a combination of theory and methods—enables us to update and inform empirical hypotheses about how the knowledge base transforms the institutional relations of an increasingly knowledge-based society. Policy implications are specified.

# Chapter One

SYSTEMS OF INNOVATION, "MODE 2", AND THE TRIPLE HELIX

In the period before the oil crises of the 1970s, that is, in the decades after World War II, social functions were deliberately organized into institutions on a one-to-one basis (Merton, 1942; Bush, 1945). Academic funding, for example, was by and large based on internal processes of peer review (Mulkay, 1976). The oil crises of the 1970s, however, made clear that advanced industrial nations could outcompete low-wage countries only on the basis of the systematic exploitation of their respective knowledge bases (e.g., Nelson & Winter, 1977 and 1982; Freeman, 1982; cf. Freeman & Soete, 1997).

The policy implications of this conclusion were not simple. Innovation is based on knowledge flowing and recombining across interfaces (Kline & Rosenberg, 1986). Knowledge flows both within and across institutional boundaries. The crossing of institutional boundaries can be expected to imply transaction costs (Williamson, 1975), but it may also generate longer-term revenues and synergies (e.g., Faulkner & Senker, 1995). The transaction costs can be considered as investments in establishing new relations of collaboration and competition. Thus, a dynamic view of a knowledge-based system can be generated in which institutional agents have to translate between short-term and longer-term optimizations using a variety of criteria.

The trade-off between transaction costs and surplus value can become visible in the changing patterns of collaboration. At the interfaces, new forms of organization can be invented. For example, the increase of co-authored papers can be used as an indicator of the increasingly networked character of research (Wagner, 2002). The newly emerging structures can be expected to reconstruct the old ones. These reconstructions may take place at different levels of the complex system of knowledge production and control.

## *1.1   The Organization of Knowledge-Based Communications*

After the second oil crisis of 1979, the techno-sciences such as biotechnology, information technologies, and new materials rapidly became the top priorities for stimulation policies at the national level in the advanced industrial countries (OECD, 1980). These "platform sciences" (Langford & Langford, 2001) are based on the assumption that rearrangements across disciplinary lines may generate competitive advantages, through synergies in the knowledge base, that can be exploited for economic development. Previous attempts for more direct mission-oriented steering of the sciences had at that time already been evaluated as less successful (e.g., Van den Daele *et al.*, 1977; Studer & Chubin, 1980).

The stimulation of university-industry relations became a second point of attention of S&T policy makers. Why had some countries been more successful than others in the technological exploitation of their knowledge base (Hauff & Scharpf, 1975; Irvine & Martin, 1984)? Why were certain sectors (e.g., chemistry, aircraft) within countries more successful than others in exploiting their respective knowledge bases (Nelson *et al.*, 1982)? Could lessons be learned from best practices across sectors, and might such practices be transferable from one national context to another?

It is far from obvious at which level one can stimulate a knowledge-based innovation system. Should one focus on optimization at the level of the institutional arrangements (e.g., Rothwell & Zegveld, 1981), or rather stimulate specific science-technologies (Leydesdorff & Gauthier, 1996; cf. Lissenburgh & Harding, 2000)? The uncertain definition of the unit of analysis for studying a knowledge-based system of innovation in terms of nations, sectors, technologies, regions, etc., brings in new players as potentially important contributors.

From the mid-1980s onwards, the European Union heavily experimented in a series of Framework Programs with policies for science, technology, and innovation. Both transnational cooperation and cooperation across sectors were systematically stimulated. Within the newly emerging context of the European Union, regions tried to promote their position as a relevant level for systematic development of the knowledge infrastructure. In the U.S.A. the national system experimented with granting rights to patent to universities (the Bayh-Dole Act of 1980), along with systematic efforts to raise the level of knowledge-intensity within industry, both at the level of the states and by stimulation programs at the level of the federal government (Etzkowitz, 1994).

How successful have these attempts been? Has a European system of innovations emerged in competition to the underlying "national" systems? To which extent have European regions (e.g., Flanders and Catalonia) been successful in establishing their own systems of innovation (Leydesdorff, Cooke, & Olazaran, 2002)? Have sectors (e.g., ICT) been developed using patterns of innovation different from those that were established in a previous cycle of industrial development (Barras, 1992)? Have patterns of collaborations across national boundaries, sectors, and disciplinary affiliations changed, and what have been the effects of these changes in terms of the quality and quantity of the respective outputs? How can systems of knowledge-based innovation be assessed in terms of their relevant outputs?

These empirical questions became even more pressing during the 1990s with the emergence of the Internet, which added a new dimension to the existing systems of innovation. The resultant global perspective makes another evaluation possible. On the one hand, the Internet was expected to increase the chances for new partners to participate in knowledge production processes by providing almost free access to information sources worldwide.

For example, South- and East-Asian countries seemed initially better equipped than European nations for moving ahead in this new era of e-business, given their specific mix of human resources and their flexibility in recomposing industrial structures and knowledge infrastructures (Freeman & Perez, 1988). On the other hand, it has been noted that the Internet tends to reproduce the stratification in the access to information and perhaps even increases the barriers of entry to markets.

How should European countries act and react? Would it be sufficient to stimulate ongoing processes of change, or should new frameworks be proposed that enable collaborating partnerships to be developed? Which criteria for the optimization should be used (e.g., national, transnational, sectoral)? Thus the stage was set for a profound reformulation of S&T policy-making in the early 1990s.

## *1.2  The science system under the condition of globalisation*

The rise of the Internet and the global dimension raised a new question in S&T policies. How do "internationalization" and "globalization" affect systems of organized knowledge production and control (Crawford, Shinn, and Sörlin 1993; Cozzens *et al.* 1990; Ziman 1994)? In a policy-oriented reflection of these developments, Gibbons *et al.* (1994) proposed to make a distinction between "Mode 2" and "Mode 1" types of the production of scientific knowledge. Whereas "Mode 1" refers to the previous shape, largely confined within institutional settings, "Mode 2" is communication driven. Knowledge can then be considered as a codification of communications.

Scientific knowledge can be contained within an institution or even an individual agent as "tacit knowledge," or it can be "published." The dimensions of knowledge in private and public arenas resonate with the arrangements of industry-government relations within political economies. The knowledge component adds a new dimension to the so-called "differential productivity growth puzzle" (Nelson & Winter, 1975) between sectors in the economy, and to the relations between public control and private appropriation of competitive advantages. The competitive advantages of nations are increasingly dependent upon scientific and technological progress (Krugman, 1996). During the 1990s, knowledge-intensity thus became a driver of the reform of the political economies.

Etzkowitz & Leydesdorff (1995) proposed to model the evolutionary dynamics of the knowledge-based economy as a "triple helix of university-industry-government relations" (cf. Leydesdorff & Etzkowitz, 1996 and 1998). According to the Triple-Helix model, three functions have to be fulfilled within a knowledge-based system of innovations: (i) wealth generation in the economy, (ii) novelty and innovation that upset the equilibrium seeking mechanisms in (semi-)market systems, and (iii) public control and private appropriation at the interfaces between economic and scientific production systems.

The specific arrangements require interfaces between the three function systems to be institutionalized as a knowledge infrastructure. However, the local stabilizations and trajectories are under pressure from global developments. The latter can be considered as the prevailing regime. In the Triple Helix model this overlay is operationalized as the communication network between the institutional partners. The knowledge-based regime of expectations guides the negotiations among the partners in the Triple Helix as an overlay system of communication, negotiations, and programming.

Under the condition of globalization, local niches can gradually be dissolved because new horizons offer other options. As the relative weights of relations in a network change by ongoing processes of collaboration, appropriation, and competition, the new balances and inbalances can be expected to generate a feedback in the knowledge infrastructure at other ends. The (sub)systems can then be expected to recombine into new solutions with degrees of success. However, knowledge flows between systems can also be expected to be stabilized and further developed within the historical institutions that have served us hitherto. In this way, institutions may survive in a changing environment. The institutional arrangements provide the stability that is necessary to access the ultra-stability of the globalized regimes (Luhmann, 2002, at p. 396).

Note that the new options are locally imported from the global level as expectations, and therefore these reconstructions are knowledge-based. Knowledge-based innovation increasingly makes the innovated systems also knowledge-based. The knowledge infrastructure is provided by networks among industries, academia, and governments. These three actors are interwoven as institutions in a network which carries the resulting knowledge base. The latter can be considered as a system of communications on top of the institutional carriers. While the institutional networks integrate, the communication systems can be expected to differentiate in terms of functions (Luhmann, 1984 and 1990). The reconstructed codifications enable both participants and observers to specify and change the systems under study inventively, that is, by proposing and codifying new combinations. This hypothesis will guide us here to map the systems under study.

A knowledge-based cultural evolution is thus envisaged which abstracts from and experiments with both the "natural" and institutional bases of the carriers at the level of the knowledge-infrastructure. The focus is on the overlay of communications. This knowledge base is meta-stabilized or globalized as an operation on the stabilizations provided by the infrastructures. The natural and institutional bases can then be considered as givens and constraints on which the knowledge-based system operates by innovating both itself and its boundary conditions.

*Co-existence, co-evolution, and lock-in*

We emphasize that a systems-theoretical approach focusing on the network level allows for a specific perspective on the interactive mechanisms between the subsystems. This perspective can be enriched with the results of institutional analyses at the lower levels. However, it adds to the latter by providing a perspective on the knowledge-based dimension that "catalyzes" the innovative processes of reorganization. This perspective, therefore, merits further exploration.

For example, the differently codified systems can develop co-evolutions when a coupling between two of them is made structural. Depending on the quality of the interaction, that is whether the interaction is supportive or competitive, one can expect co-existence and/or selection. Co-existence is the outcome of continuously generated stabilities between counteracting mechanisms within the overall system. The co-evolution then generates a process of "mutual shaping" between the co-evolving systems. When a third dynamic is added to such a co-evolutionary model, previous arrangements can be dissolved at a global level. The system can therefore be expected to shape stable trajectories and global regimes endogenously (Leydesdorff & Van den Besselaar, 1998).

The possible outcomes of the interplay among three subdynamics can be richer than in the case of two dynamics, as a new quality of interactions is introduced. Chaotic trajectories are also possible at this level. While there is no longer an essential solution or harmony in such "trialectics," one can expect the "endless frontier" (Bush, 1945) to be replaced with an "endless transition" (Etzkowitz & Leydesdorff, 2000). The production of both new partners for interactions and new types of interactions is an endogenous feature of such complex dynamics.

Each new dimension raises the number of possible realizations exponentially.[1] New institutional forms, for example, can serve as boundary-spanning mechanisms that enable the participants to specify new variations. These processes can be modeled. Although a prediction of specific variants with the help of these models remains principally impossible—because the model abstracts from the substantive content—the boundary conditions for successful variants can be tested in simulations (Ebeling & Scharnhorst, 2000). The systems of innovation can be expected to compete in their uphill search for new solutions and stabilizations (Kauffman, 1993; Frenken, 2000).

For example, trajectories can be formed by chance processes at interfaces when technologies are "locked-in" within industries (e.g., the QWERTY keyboard; David, 1985). Alternatively, scientific expertise and state apparatuses may begin to co-evolve such as in the energy and the health sectors. The state and industry can also become "locked-in" like in the former Soviet-Union. Innovation policies have to vary in terms of which "lock-ins" (between co-evolving subsystems) are prevalent, and on the assessment of how these patterns can be systematically disturbed by a third dynamic. For example, the market mechanism can reintroduce flexibilities in the case of a bureaucratic lock-in, whereas, in the case of a technological lock-in, government interventions may be needed to break monopolistic tendencies. Thus the optimization of policies becomes increasingly *dependent* on the evolutionary assessment of the knowledge-based system.

This dependency relationship tends to invert the cause-effect relationship in political steering processes. The room available for steering is increasingly determined by the systems to be steered. However, the self-organization of the latter at the global level can be reflected and then made the subject of informed and knowledge-based policy-making. Whereas the lock-in phenomena can make the system robust against steering for long periods of time, this process of stabilization is under permanent pressure from an evolutionary perspective. The global perspective destabilizes local stabilizations. Systemic innovations are possible because of this destabilization. However, destabilization can also lead to a collapse. The challenge for a complex system is to balance between the ability to innovate and stability. Innovation policies can then reflect and/or disturb this balance, but without further development of their own knowledge base the policy makers can be outcompeted by the knowledge bases of the (sub)systems to be steered.

---

[1] Two dice provide $6^2$ (=36) possible combinations, while three dice provide $6^3$ (= 216) combinations.

*Globalization, stabilization, and reflexivity*

A global regime results from closer interactions among relatively autonomous subsystems, for example, in terms of networks of university-industry-government relations. The global regime is propelled as a complex dynamic among differently coded communication systems (e.g., the economy, science, and policy-making). The network overlay emerges as a new unit of evolution. When this structural innovation can be temporarily stabilized, it may begin to coevolve with the subdynamics upon which it builds.

Given the selection pressure of the new dimension, old institutional arrangements may survive, but will probably have to adapt their function, as well as their form, to the new environment. The hypothesis of a "global agent" can be formulated as the expectation of change because of the selection pressure on institutional arrangements. The global agent, however, remains a network function and consequently operates as a regime of uncertain expectations. It is not a steering agent with a positive agenda, but a global regime that exerts selection pressure by being pending.

One should not reify this "global agent" as a metabiology or a supersystem. The various systems of expectations interact and produce an overlay *within* the system of interactions. This overlay globalizes the system by making other representations available compared to those that could already be envisaged from the previously available perspectives. These recombinations can then be attributed to a next-order or "global" system, but their possibility is only a result of an internal dynamic that is added to the system as its "globalization." This globalization can be entertained reflexively therefore enriching the system. It provides a future-oriented knowledge-base that innovates the historical systems with hindsight. The ability to innovate is based on inventing new codifications by reflexively rearranging at the borders.

The dynamics of science and technology have induced a reflexive turn in other social systems. The effects of being increasingly knowledge-based have first been reflected in science and technology studies (e.g., Whitley, 1984). The "reflexive turn" in these studies (Woolgar & Ashmore, 1988) implied that the idea of a single and universal yardstick—as searched for in the philosophy of science (e.g., Popper, 1935)—had to be given up in favour of codes that are continuously constructed and reconstructed. Unlike universal standards, asymmetry can be expected to prevail in exchange relations among systems and subsystems (Gilbert & Mulkay, 1984) because the systems exchange on the basis that they have different substances in stock.

For example, the political system is initially interested in results from the science system that inform decision making and policies without being unduly burdened with the overwhelming uncertainties that are intrinsic to scientific inferencing. However, within the science system these uncertainties may raise new, and possibly fundamental research questions (Beck, 1986). Similarly, the science system can develop reflexively in relation to problems arising in industrial contexts (Rosenberg, 1976). Often new opportunities to patent arise unexpectedly within the research process. In other (e.g., industrial) contexts, scientific progress can sometimes be considered as an unintended side-product, with the intended focus being on problem-solution.

Rosenberg (1982) raised the question "How exogenous is science?" Ever since, the non-linear dynamics in the science/technology interactions has taken the lead in the research program of science, technology, *and innovation* studies (STI)—as we have now began to call this field of expertise (Wouters *et al.*, 1999). Whereas the sciences are developing along historical lines, innovation reorganizes the systems on which it builds at the interfaces. This continuous reorganization under the pressure of competitive innovations has been institutionalized in advanced industrial systems since the scientific-technical revolution of the period 1870-1910 (Braverman, 1974). Since then, further development of technologies takes place at the interfaces of the sciences, the economy, and the useful arts (Noble, 1977). In a later part of this study, we will distinguish between science indicators in terms of scientific communication, technology indicators in terms of patents that map technological inventions, and innovation indicators that may also map market introduction, for example, at the Internet. We propose to recombine these indicators reflexively in a program of innovation studies.

## 1.3 *From theories to empirical evaluations*

*Units of analysis versus units of operation*

The analytical models provide us with heuristics for the empirical research. However. the knowledge component of systems cannot directly be observed as organized knowledge acts as a different operator to the observable instantiations that it changes (Giddens, 1984). The knowledge systems studied operate dynamically. A representation provides us only with a picture of the footprints of previous communications.

Scientometric and webometric indicators trace functions of communication. Functions can be attributed to institutions. For example, publications and citations span networks of communication, but one can also use them for ranking institutions in terms of their productivity. Note that we wish to alter the research focus: one can rank scientists, that is knowledge carriers, but scientific communications develop at the network level. Networks tend to develop in different directions with different qualities. Over time, the network dynamics may redefine what has been a significant contribution and in which respect. The dimensions (functions) of the networks can be considered as increasingly orthogonal. Thus, the networks group the communications instead of ranking them.

The significance of a contribution is not an inherent property of a contribution, but a construct "in the eye of the beholder" (Latour, 1987; Leydesdorff & Amsterdamska, 1990). The windows for studying subjects as intangible as knowledge production and communication, have to be carefully reflected as the order of communications is not "naturally" given. We are constructing second-order constructs about knowledge-based constructs.

Some authors have proposed the consideration of "national systems of innovation" as the appropriate unit of analysis for innovation studies (Lundvall, 1992; Nelson, 1993). The choice for a national perspective allows for a direct link to the possibilities and limitations of policy making by national governments. Furthermore, it enables the researcher to use national statistics (Lundvall, 1988). However, from a reflexive angle, each communality or dimension can be considered as a construct that can be more or less codified.

For example, the notion of a national identity may be changing from a European perspective. The construction of a regional identity, for instance, has resounded in some regions because of linguistic differences, but in others, such as in France, regional authorities have been shaped in order to accommodate to European policies and harmonization. In other words, the units of analysis and the systems of reference can be analytically considered as constructs that then tend to shape the analysis. The windows that we use provide us with a metaphor that can easily turn into a bias or a metonym. What can be considered as highly relevant from one perspective, may be contextual from another. The categories in which science, technology, and innovation studies reconstruct bodies of knowledge have to remain hypotheses! It is precisely as hypotheses that the concepts invite to proceed to the operationalization and measurement.

*The time dimension*

In contrast to a historical build up, the evolutionary dynamic continuously operates in the present and with hindsight, that is upon the instantiations of the systems under study as its basis. Thus the global dimension tends to invert the historical time axis in the analysis. Whereas growth-rates, for example, are usually expressed with reference to a previous moment in time and time series are standardized with reference to a historical moment (e.g., "1990 = 100"), **the evolutionary perspective is policy relevant because the analyst can take the present as the system of reference**. The present state contains the analytical reconstructions as representations of its past. Thus the evolutionary analysis provides information from which one can develop options with a greater or lesser degree of success, without prescribing future behaviour in any sense.

Hitherto, statisticians have had an inclination to build on their resources using a historical perspective. Sociologists interested in history may then be able to use these materials as illustrations in support of their narratives. However, the focus on knowledge-intensive developments requires us to take a reflexive turn towards the data gathering process, both in the quantitative and in the qualitative domain. The program of innovation studies is anti-positivistic, as one begins with expectations instead of the observable "facts." The facts mean different things at different sides of an interface.

From an evolutionary perspective on cultural phenomena such as science and technology, the analyst first specifies which assumptions went into the data collection and whether these assumptions are still valid when, at a later stage, one raises new questions from the evolving science and technology policy agenda. For example, when studying the development of journal structures in "biotechnology," one has several options. If one fixes the journal set *ex ante*, one observes the development of "biotechnology" as conceptualized at the beginning of the data collection (e.g., in 1985). If one defines the journal set dynamically, one studies the changing meaning of "biotechnology" in relation to other journals. If one fixes the journal set *ex post*, one refers to the understanding at the later moment in time (e.g., in 2003).

The analysis of the historical strengths and weaknesses of a research portfolio does not itself suggest that one should "pick the winners" (Irvine & Martin, 1984) in order to strengthen one's case globally, that is at the system's level. The "winners" may have been yesterday's winners and one may have other reasons to strengthen the hitherto relatively weak groupings or clusters (Porter, 1990). The empirical analysis informs us about the contingencies that can be expected. However, as the dynamics are complex, unintended consequences and unforeseen externalities can also be expected. The formative evaluation during the process provides us with signals that can then be made the subject of more systematic analysis.

*Operationalization, data, and context*

A crucial step in proceeding from the formulation of analytical hypotheses to the collection of empirical data is implied by the concept of operationalization. How can one move from the analysis to the indication of the importance of the concepts in a social reality? How can a reflexive analyst make a convincing argument when the notion of a system of reference can always be deconstructed, and the time line can be inverted in terms of what the historical account means for the present?

As systems that contain knowledge should not be considered as given or immediately available for observation, one has to specify them analytically before they can be indicated or measured. In the end, the quantitative analysis depends on the qualitative hypotheses. For example, one can raise the question of whether "Mode 2" has prevailed in the production of scientific knowledge. What would count as a demonstration of this prevalence and what would count as a counterargument? Can instances be specified in which one would also be able to observe processes of transition between the two modes? What should one measure in which instances and why?

While the qualitative analyst reduces the complexity by taking a perspective, quantitative analysis allows for the question of the extent to which a perspective highlights a relevant dimension. How much "Mode 2" is in the development of biotechnology in Germany as compared with the development of biotechnology in the United States? A policy analyst is always able to indicate contingency, similarities and differences, continuities and changes, but the quantitative analysis requires that these categories are specified as ex ante hypotheses in order that the expectations can be updated by the research efforts. The empirical research should enable us to specify the percentage of the variation that can be explained using one theoretical model or another.

Whether "Mode 2" is "old wine in new bottles" (Weingart, 1997) or new wine in old bottles depends on the definitions of the bottles and the wines, and the processes of change that are analytically outlined in the research design. The definitions with reference to a knowledge-intensive system are knowledge-intensive themselves (Nowotny *et al.*, 2001). The observations and indicators are also knowledge-intensive, as one can no longer assume that the data is readily at hand. The overwhelming availability of information in a knowledge-based society makes it necessary to reflect on the selection of data from a theoretical perspective.

Established indicators were time-stamped in a previous period and historical evidence can retrospectively be recognized as anecdotical. Systematic data collection, however, requires standards. The matching between the analytically relevant questions and the institutionalized routines asks for an informed trade-off between considerations of a potentially very different nature. How does one define a baseline? How does one normalize? What is/are the relevant system(s) of reference? Scientometric indicators cannot simply be "applied" in another context without generating terrible confusion. Scientometrics is a research effort in its own right, since the indicators have to be reflected.

For example, the debate of "the decline of British science" (Irvine *et al.*, 1985) as measured in terms of publication performance data was paradoxically possible because "British science" had been relatively stable (Braun *et al.*, 1991; Leydesdorff, 1991; Martin, 1991 and 1994). Thus, methodological decisions as to whether the analysts used an *ex ante* fixed journal set in order to make comparisons along the time axis possible or followed the development of the dynamics of the journal sets included in the *Science Citation Index* had an impact on whether one measured decline or not.

The further decision to attribute each publication with a British address to the UK with a full point or only pro ratio of the number of corporate addresses in internationally coauthored publications (so-called "fractional" versus "integer" counting) includes an effect of internationalization on the performance measurement that can be expected to differ from nation to nation. Integer counting, however, leads to problems in the normalization since the sum-total does no longer add to hundred percent (Leydesdorff, 1988; Anderson *et al.*, 1988; Braun *et al.*, 1989).

These methodological problems reflect decisions that have to be taken on the basis of arguments. The theoretical grounds can be made relevant for the scientometric enterprise if they can be formulated as hypotheses that are operationalized reflexively before one is able to draw conclusions. The selection of data is necessarily specific and this specificity has to be reasoned.

## 1.4 Statistical analysis, models, and simulations

*Multidimensionality and the reduction of uncertainty*

In principle, data inform the hypotheses, but not by themselves and not necessarily in the positive. Data may also confuse us; particularly when they are so abundantly available as nowadays. A variety of representations is always possible and this problem is further aggravated when databases are no longer substantively codified and dedicated, but when algorithmic search engines and meta-crawlers become widely available. The data provide us first with variation and therefore uncertainty, and the perspectives on the data may also be at variance.

Specification of a reflexive perspective reduces the uncertainty. This can also be expressed formally by using probability theory. First, a probability distribution of a variable can be hypothesized. Then, each further specification can be considered as an additional condition to this probability distribution. The third law of probability calculus specifies that the likelihood of two probabilities (A and B) together is equal to the likelihood of A given B times the likelihood of B. Or in formal language:

$$p(A \text{ and } B) = p(B) \cdot p(A|B)$$

Since all probabilities are smaller or equal to one, the uncertainty in the distribution *A* is reduced by our knowledge of the distribution *B*, unless the two distributions are completely independent. Whether the distributions can be considered as independent or not can be tested using significance testing (e.g., chi-square). If there is significant dependency, one is allowed to conclude to a reduction of the uncertainty in the prediction. Therefore, multidimensionality not only enriches the complexity of the problem. It also provides a way to reduce complexity. The specification of conditionality (that is, "contexts") reduces the uncertainty in the system under study (the "text"), unless the context was irrelevant.

The dependence or independence of two variables can be detected by analyzing their co-variation. If the two variables represent two different systems (or subsystems), these systems determine each other through this window of co-variation or "mutual information," but otherwise they only condition one another. Thus, the language of the quantitative analyst replaces expressions like "enabling and constraining" (Giddens, 1984) with concepts of determination, reduction of uncertainty, and conditionality. The "mutual information" provides the systems with windows upon each other. A co-variation when repeated over time may develop into a co-evolution between two systems.

*History, co-evolution, and the emergence of new systemness*

In many cases, one can build on existing definitions of systems—like "the research system"or "the patent system"—but in the case of knowledge-based systems one may also be interested in "emerging systemness". Emergence can only be analyzed by observing the interaction among systems over time. From this perspective one can analyze the evolution of each system along a trajectory. However, one can also focus on the interaction and potential co-evolution between systems? Did the co-evolutions lead to a new system or did the (intended? questioned?) outcome fail to be realized?

If there is co-variation over time and then also co-evolution, one can expect the emergence of a degree of systemic development. However, the question for the evaluation remains whether at a certain moment in time (e.g., today), systemness is prevailing over historical variation or not. These two dimensions—historical variation and systemness in the present—can be considered as analytically independent.

The subsystems develop historical variation along potentially different trajectories, but the next-order system selects by weighing among the trajectories so that it can maintain its system's order.[2] When a system evolves over time, one can ask how the state of the system at time $t$ depends on the state of the system at a previous moment ($t - 1$). This relationship can also be turned around. Then, one asks how the state of the system at time $t$ determines the state of the system in the future one time step ahead (that is, at $t + 1$). The Markov property states that the best prediction of the next stage of a system is its current state (Yablonsky, 1986). "Markov systems" have no long-term memory about historical orders at lower levels because the system is able to reorganize in the present using the relative weights of the various subsystems that it recombines.

In science and technology policy the appearance of a next-order system is often a question more interesting than the expectation of stable development along a trajectory (Allen, 1994). For example, in the case of Europeanization one can raise the question of whether a European dimension of the publication system in a specific field can be discerned. One can test this hypothesis by measuring the publication output for each of the individual nations historically and then make a prediction on the basis of the respective time series that can be compared with a prediction based on the assumption of emerging systemness.

---

[2] Systems with non-linear interactions exhibit additionally the capacity to develop different scenarios which may branch in time like a tree. This has also been called path-dependent development. Once, a trajectory is chosen, the system is bounded to a path. One could say that a system once codified by this "lock-in" becomes a potential candidate for a next-order selection.

| Country   | 1990        |  |  |  | .... | 2001        | 2002        | 2003 |
|-----------|-------------|--|--|--|------|-------------|-------------|------|
| Country A | $a_{1990}$  |  |  |  |      | $a_{2001}$  | $a_{2002}$  | ?    |
| Country B |             |  |  |  |      |             | $b_{2002}$  | ?    |
| ...       |             |  |  |  |      |             |             | ?    |
| Country N |             |  |  |  |      |             |             | ?    |

**Figure 1.1**
*Time series of rows versus systemness over the columns*

For example (using Figure 1.1), one can make a historical prediction of the publication performance of Country A in the year 2003 on the basis of the values of the indicator $a_{1990}$ to $a_{2001}$, and similarly for country B, etc. The alternative prediction would be that systemness has grown among these (European) nations and that the European dimension would prevail. In that case, the situation in 2002 would provide us with a snapshot along the column dimension of Figure 1.1 of how far this system has developed. The best prediction of the situation in 2003 would then be based on the Markov assumption that the current state (2002) would be maintained and reproduced as a distribution in the next year. As soon as one is able to measure the publication volume for the year 2003, the two predictions can be compared.

One cannot reject a hypothesis on the basis of a single measurement point, but the principle of testing two hypotheses against each other may be clear. The two predictions above are analytically independent since based on the rows and columns of the matrix, respectively. Therefore, the predicted values are different and they can be compared with the measurement results. One can also hypothesize that the observed values are to a certain degree (e.g., 30%) predicted by the one hypothesis and to a complementary degree (70%) by the other. Thus, one is able to specify the percentage of the variation that can be explained by using one assumption or another.

Using these methods, a European publication system could, for example, not yet been discerned in terms of the publication data included in the *Science Citation Index* (Leydesdorff, 2000a). In another study, Leydesdorff & Oomes (1999) were able to show how the emerging European Monetary System (EMS) affected national systems in the monetary and economic domains, respectively, during the period 1985-1995.

Let us now proceed by generalizing the analytical independence of the prediction based on the rows versus the columns of a matrix to all scientometric tables and spreadsheets that are thus designed. The two dimensions of an (asymmetrical) matrix refer to different dimensions of the system under study or—in other words—different systems of reference. For example, the scientometrician can count word-occurrences in documents. The documents are then considered as the cases and the words as the variables. The words provide us with an indicator of the intellectual organization of the documents, whereas the documents can also be grouped in terms of their institutional (e.g., national) addresses. Thus, a matrix of words versus documents provides us with information in the communicative dimension of the intellectual exchange and the dimension of the institutional organization. These two dimensions (subsystems) are coupled by the research design when specifying a hypothesis that can be tested.[3]

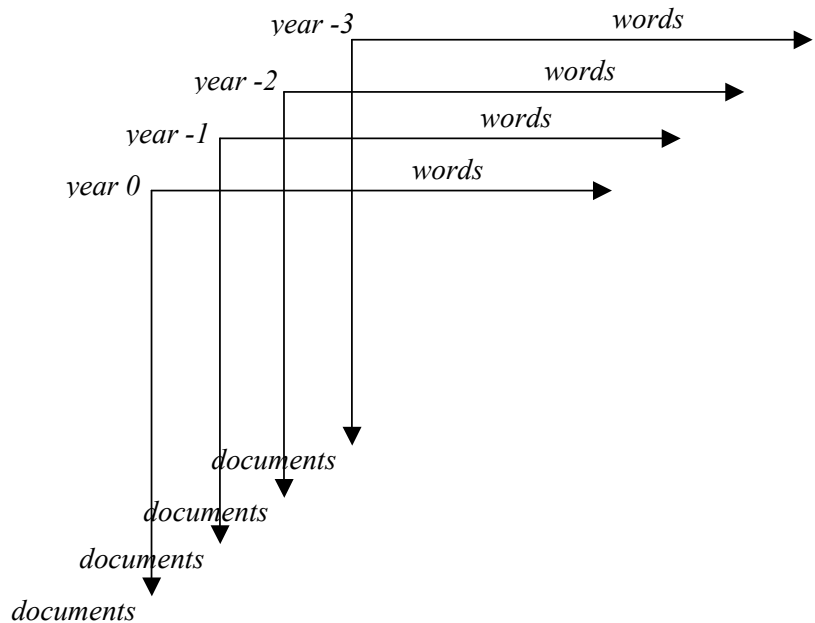

**Figure 1.2**
*Matrices of words versus documents in a time series*

When one constructs such a matrix of words versus documents for each year in a series of years, one can place these matrices behind each other and then one would obtain a cube of information. Along the time axis one is able to ask whether the words (indicating the intellectual organization) have grown into a system or whether the institutes (represented as aggregates of documents) have rearranged their relations. In each of these dimensions one can additionally ask the question to which extent systemness has become prevailing over historical variation (or not). Finally, one can also raise questions of whether the co-variation in the matrices for each year has become increasingly systemic over time.

In Leydesdorff & Heimeriks (2001) this methodology was applied on publication data in the field of biotechnology. The conclusion was that the American science system had been more self-organizing than the European system in the intellectual dimensions as measured in terms of coherent word usages, perhaps because of the prevalence of national tendencies in word usage among the European nations. The European system, however, could be shown to have effects in the institutional dimension (Lewison & Cunningham, 1991).

---

[3] In the case of a symmetrical co-occurrence matrix part of the information has been discarded by multiplying the original matrix with its transposed.

*Models and simulations*

The focus of our discussion has hitherto been on how self-organization, co-evolution, and systemness can be analyzed by measuring sets of variables at different moments in time. One level deeper, one can ask which dynamics can be expected to lead to the temporal and structural changes that one observes. Modeling and simulation focus on developing explanations for the observable patterns.

Modeling efforts and simulation studies can be retrieved from the literature in science and technology studies, but these studies have been scarce (Goffman, 1966; Nowakowska, 1984; Kochen, 1983; Ebeling & Scharnhorst, 1986; Wagner-Döbler & Berg, 1993). Perhaps, models of stochastic processes have been an exception to this rule (e.g., Price 1976; Egghe & Rousseau, 1990). In evolutionary economics, however, models have been extensively used as tools for explaining technological developments (cf. Fisher & Pry, 1971; Silverberg, 1984; Arthur, 1989).

In the 1990s a new type of model (so-called "agent-based" models) gave modeling and simulation a further impetus in science and technology studies (Axtell & Epstein, 1996). Because these models start from rules for individual behavior, they are suitable to model social processes that are based on assumptions about agency (Gilbert & Troitzsch, 1999). Processes in the sciences like citation and publication behavior, but also the diffusion of technologies and the appearance of the web, have since then been modeled from this perspective (Bruckner et al., 1990; Gilbert, 1997; Leydesdorff, 2001c; Scharnhorst, 1998 and 2001; Boudourides & Antypas, 2002)

The modeling approach opens a different perspective on indicator research. On the basis of the conceptual framework of a model, new measurement requirements can be specified. Whereas the scientometric measurement has focused on the historical cases that actually occurred, a model first specifies a realm of possible events. The actual events can then be considered as instantiations. The instantiations are determined by parameter values. The parameter values have to be estimated on the basis of observations.

Let us elaborate using an example. The growth of scientific disciplines can be modeled as a process of competition. The growth curve of a single field can be considered as an outcome of the interplay between different fields (Bruckner *et al.*, 1990). In terms of population dynamics, the scientific fields are then defined as the units of evolution. These units, however, are not self-standing agents, but collectives of individual agencies. The higher-level units can be considered as patterns of coherent behaviour of scientists who have grouped together at a lower level in "invisible colleges".

The interactions between different fields are caused by the underlying choices of the scientists working on certain topics and not on others. Scientists may move between fields by changing their research focus. By assuming the fields as the units of evolution in science, this movement can then be modelled as a competition between the fields on the available scientists as human resources (Gilbert 1997). The scientists, however, are steered both from the control level of the field and, for example, in terms of available funding. Thus, the model develops at two levels at the same time and with feedback loops between these levels.

In such a model, various processes at the micro level can be distinguished. For example, the educational process of scientists can be considered as an entry process to a field, the mobility between fields as an exchange process, and the career ending of scientific activities in a field as an exit process. These various processes have to be weighted by using parameters. The estimation of these parameters, however, can only be based on measurements.

Confronted with the task to validate parameters, one becomes aware that this information is not gathered by scientometric indicator research in a way that can easily be transformed into the parameter values for simulations. The indicators tend to focus on specific processes, but not on the interaction terms between different processes. For example, one can easily find data about the education of scientists in different disciplines, but data about the number of "newcomers" in scientific specialties are far more difficult to retrieve. Migration pattern of scientists between specialties are seldom analysed and then not easily connected to the growth of specialties (e.g., Mullins, 1972; Mulkay, 1977).

In the case of scientometric indicators, longitudinal data are often difficult to obtain since the focus in scientometrics has been on what can be called "comparative statics." How has the situation changed since a previous moment in time in terms of the observable data? Indicators then provide snapshots for different moments in time. The dynamic analysis is different from "comparative statics" since the latter approach does not aspire to analyze the processes underlying the observable changes.

In summary, modelling efforts create a demand for new or refined measurement instruments. The estimation of parameters is oriented to the underlying processes at a micro-level, while current indicators tend to focus on the observable phenomena at a macro-level. Note that simulation models can also be used to produce "virtual" indicators. Hypotheses about underlying processes of change in knowledge production can be turned into specific subroutines of a dynamic model. Simulations produce quantitative output that can be compared with what happened. The comparison between the statistical and the "virtual" indicators can be further analysed, for example, for explaining what caused the difference. By comparing "virtual" indicators with empirical measurements hypotheses behind a model can sometimes be tested.

How can one compare the historically observed values with the evolutionary expected ones? In this study, we do not elaborate further on modelling, but we focus on the measurement. However, we wish to emphasize that modelling and simulation can be understood as a part of indicator research as both research programs are interested in the quantification of the description and then also the explanation. The theoretical challenge consists of the creation of a link between the quantities of the models and the quantities in the empirical observations. The selections both in terms of relevant (empirical) data and in terms of assumptions in the simulation models have to be guided theoretically if we wish to relate these two domains as research programs in innovation studies.

## *1.5 Conclusion*

Quantitative indicator research develops on the edge of evolutionary modeling and historical observations. The historical observations of communication can be expected to contain uncertainty because the very concept of communication implies an exchange. Thus, communication is distributed by its very nature. From this perspective, qualitative theorizing contributes by providing hypotheses, i.e., uncertain expectations. The challenge is to relate the specified expectations to observable data. The hypotheses can first be used to distinguish structural uncertainty from random fluctuations, error, and noise. The simulation model adds the interaction of different processes to the hypotheses and it allows for experimentation with possible scenarios of systems development.

The space of possible future developments can only be accessed algorithmically. Once a set of variables has been defined, the algorithm describes the temporal changes of these variables by looking at the fluxes (dx/dt). To gain understanding, however, the analyst uses geometrical metaphors based on times series of variables *or* the analysis of the multi-variate complexity as instantiations in the present. One would overstress one's linguistic capacities by describing changes in the values and the meaning of variables in the same pass. Once the parameters are chosen for the representation, one is bound by a set of geometrical constraints on the representation. The qualitative appreciation in the narrative can therefore be considered as generating a metaphor or window on the complexities under study.

If one tries to describe both change in the meaning of the variables and the value of the variables using a single design, the comprehension tends to become vague and confused. Luhmann, for example, invoked in such instances the metaphor of a "paradox:" The algorithmic system can be expected to be more complex than the geometrical metaphor ("picture") stabilized in a discourse (Hesse, 1988). The discourse contains a perspective that can only be changed discursively. An additional change in the meaning of the variables can then be formulated as a dynamic problem. The algorithmic formulation increases the complexity to the extent that different perspectives can be entertained as competing for the explanation.

The results of this process can discursively be appreciated as the update of the hypothesis. The *ex post* picture can be different from the *ex ante* one. The *representation* is then "translated" (Callon *et al*., 1986). However, does this imply that the *represented* system is also changed? Are we able to distinguish dynamically the bias caused by our representations from the changes in the represented systems? In our opinion, quantification of these concurrent processes of change is the major research effort of quantitative studies in science, technology, and innovation studies following the reflexive turn in science, technology, and innovation studies.

The set of variables or, in other words the definition of the system, fixes an axis for the comprehension, but the codification of this system also generates a "blind spot."

Empirically, the problem of a system with potentially changing taxonomies re-appears in the choice of the units of analysis. However, can one change the unit of analysis "on the fly"? The algorithmic approach enabled us above to change to the specification of a unit of operation as different from a unit of analysis. The focus on innovation studies makes this reformulation unavoidable because innovation can only be defined as a unit of operation at an interface.

In our opinion, this change of perspective to an algorithmic approach—entailing the appreciation of narratives as heuristics—enables us to solve some of the outstanding problems in the scientometrics program. For example, the need for a historical baseline was signaled early in the scientometric enterprise (Studer & Chubin, 1980). Narin (1976) proposed to work with an *ex ante* fixed journal set as an analytical tool in order to make comparisons possible among time series data. As noted, this decision contributed to artifacts in scientometric representations of "the decline of science in the U.K." during the 1980s.

Collins (1985) raised the question about the appropriate unit of analysis for science policy evaluation. He argued against an institutional delineation of units of analysis in order to compare "like with like" (Martin & Irvine, 1983). Scientific developments, however, cannot be equated with the development of institutional units nor with fixed journal sets. The focus on flows of communication makes it necessary first to specify *what* the hypothesized (since not so easily observable) system of communications is communicating when it operates. The specification of the unit of operation extends the analysis with a specification in the relevant time dimension. Only after the hypothetical specification of the "what" of the communication, can one address the question of "how" this communication can be envisaged. The specification of the "how" of the operation then induces the specification of an indicator.

For example, on the basis of the assumption that scientific specialties are developed in terms of knowledge contributions, one can ask how knowledge is contributed. Scientific articles then become a prime candidate for measurement. How are scientific articles related? Co-words among titles and citations can then be understood as indicators of the hypothesized exchange processes. The aggregation of citations (or other scientometric indicators) enables us to map the sciences under study at certain moments in time. This methodology can be contrasted with the use of citations (and other indicators) for the reconstruction of a hypothesized development over time.

In other words, the theoretical specification constructs the (hypothetical) systems under study. For example, using a journal set provides us with a focus on the scientific publication system. Using patent data provides us with a focus on technological inventions. These two systems are differently codified and therefore can be expected to exhibit different dynamics. The study of scientific citations in patent literature, and vice versa, provides us with a focus on the interface between these two literatures, that is, patents and publications. However, there is no necessary relation between the two types of data. On the contrary, several studies (Narin & Noma, 1985; Narin & Olivastro, 1992; Blauwhof, 1995; Schmoch, 1997; Meyer, 2000a,b; cf. Grupp & Schmoch, 1999) have noted the narrowness of the window of communication between these two systems.

Others have focused on institutional relations between addresses in patent literature and scientific publications, but also, in this dimension, differentiation may prevail above integration (e.g., Riba-Vilanova & Leydesdorff, 2000). The local integration communicates among systems that first have to be specified analytically. The expectation is that the communications can locally be integrated because they are differentiated in other dimensions.

The evolutionary perspective of innovation studies makes it furthermore necessary to delineate the systems of reference from the perspective of hindsight. The hindsight approach generates a relation with future-oriented policy perspectives as one informs the reader with reference to the present state of the systems under study. For example, what we understand as "biotechnology" nowadays is something completely different from what governments wanted to stimulate in the 1980s (Nederhof, 1988). Analogously, what industries subsume under "biotechnology" as a category at present, is different from the definition of "biotechnology" by research councils. A modern society has many facets and is therefore differentiated in terms of its coordination mechanisms, codifications, and media of communication.

# Chapter Two

CAN THE NEW MODE OF KNOWLEDGE PRODUCTION BE MEASURED?

We have argued hat a fundamental reformulation of the problems of Science, Technology, and Innovation Policies became urgent during the 1990s because the following developments reinforced each other:

(1) The emergence, spread, and convergence of technological and communications paradigms such as the computer, mobile telephony, and the Internet; interaction itself has become more extensive among organizations, multi-layered, and therefore relatively more important than the elaboration of perspectives within the walls of one's own institution based on routines and tacit knowledge;

(2) The interconnection between the laboratory of knowledge-production and users of research—at various levels—exemplified by the rapid growth of industry-university centers in which firms and academic researchers jointly set priorities; technology transfer agencies within both universities and firms that negotiate with each other and move technologies in both directions;

(3) The consequent transition from vertical to lateral and multi-media modes of coordination, represented by the emergence of networks, on the one hand, and the pressure to shrink bureaucratic layers, on the other.

The authors of the "Mode 2" thesis (Gibbons *et al*., 1994) argued that this new configuration has led to a dedifferentiation of the relations between science, technology, and society. Internal codification mechanisms (like "truth-finding") were discarded as an "objectivity trap" (Nowotny *et al.*, 2001, at pp. 115 ff.). The epistemological core of science was declared not to be only uncertain, but therefore (!) completely empty. From this perspective, all scientific and technical communication boils down to communication that can be equated and compared with other communication from the perspective of science, technology, and innovation policies.

In our opinion, the study of communication can be guided by available theories of communication. Two theories are then particularly important: Luhmann's sociological theory of communication (Luhmann, 1984 and 1990; Leydesdorff, 2001a) and the mathematical theory of communication (Shannon, 1948; Theil, 1972; Leydesdorff, 1995). The crucial point becomes how to relate these two theories with different (epistemological) statuses, so that the quantitative measurement enables us to update and inform the hypotheses based on substantive theorizing.

## 2.1 *The reflection of boundary-spanning mechanisms*

University-Industry-Government relations can be considered as a boundary spanning mechanism in the knowledge infrastructure of societies. The operation of boundary spanning mechanisms indicates that the transaction costs can be balanced by the expected surplus value of the collaboration. International coauthorship relations, for example, provide another boundary spanning mechanism, but across national boundaries. Publications can also be related intellectually, for example, by being published in the same or similar journals.

The *Science Citation Index* 2000 provided us with information about 3745 journals in which articles are published, usually including institutional and national identifiers alongside author names. This data enables us to study coauthorship relations that cross institutional boundaries. The 1,432,401 institutional and corporate addresses contained in this data set were attributed with the categories "university," "industry," and "government" by using an automated routine. We were thus able to classify 86.6% of the addresses in these three categories. The identified affiliations refer to 93.3% of the 778,446 records of unique documents in the database.[4]

Once categorized, this data can be analyzed using the statistics of mutual information. The mutual information (or transmission) differs from co-variation, co-occurrence measures or correlation analysis because this measure is also defined in three dimensions.[5] When the three subsystems (university, industry, government) are completely uncoupled, the mutual information vanishes ($T_{U-I-G} = 0$). When the three dynamics are mainly coupled by sharing a communality in the variation (e.g., in the case of a hierarchical (e.g., étatist) regime or perhaps in corporatist arrangements), the value of this transmission is positive (Figure 2.1). However, when the three domains are liberally coupled through uncoordinated bi-lateral relations, this indicator can also become negative (Figure 2.2). Thus, the indicator provides us with a measure for the state of a Triple Helix system whenever the relevant relations can be counted.

---

[4] These documents were written by 3,060,436 authors so that on average each document contains two addresses and four co-author names.

[5] The transmission in three dimensions (x, y, z) can be defined as follows (Abramson, 1963, at p. 129):

$$T(xyz) = \Sigma_{xyz} P(xyz) \log \{[P(xy).P(xz).P(yz)] / [P(x).P(y).P(z).P(xyz)]\}$$

Or in terms of the Shannon notation:

$$T(xyz) = H(x) + H(y) + H(z) - H(xy) - H(yz) - H(xz) + H(xyz)$$

In the first formulation, P(x) stands for the probability of an event x and P(xy) for the probability that x and y occur together, etc. These probabilities can be measured by counting frequencies of (cooccurences) of events as will be shown in the empirical examples below.

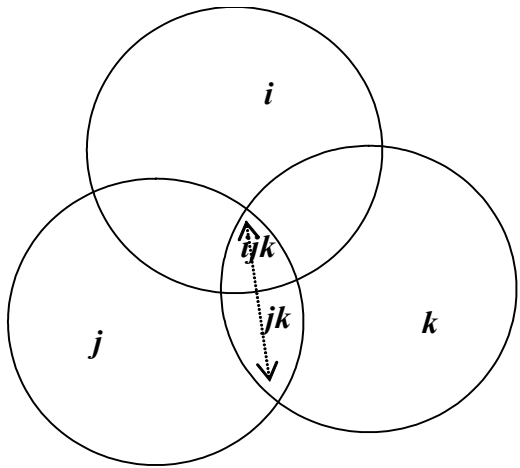 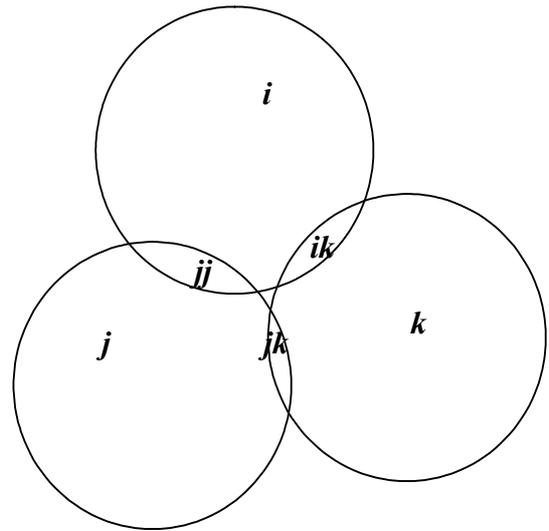

**Figure 2.1**
*Three subsystems with a center of coordination*

**Figure 2.2**
*Three subsystems without center of integration*

Conceptually, the potential generation of a negative entropy corresponds with the idea of complexity that is contained or "self-organized" in a network of relations that lacks central coordination. The system then propels itself in an evolutionary mode (Figure 2.3). The reduction of the uncertainty by this negative transmission is a result of the network structure of bi-lateral relations. Note that the mutual information in two dimensions contributes negatively to the uncertainty that prevails, while the three-dimensional overlap increases the local entropy.

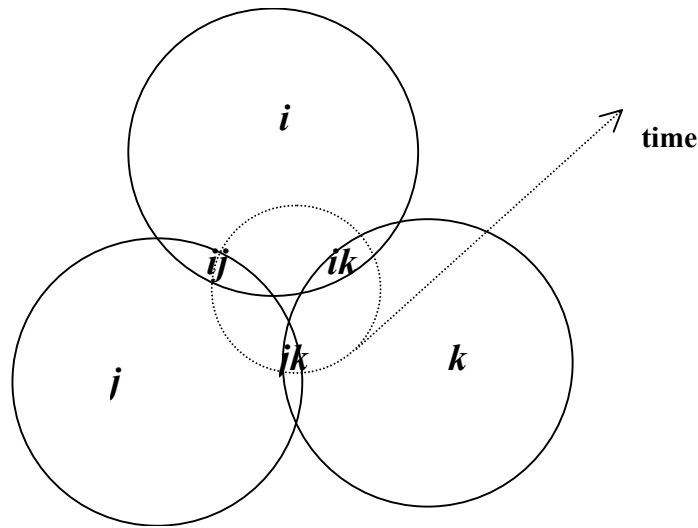

**Figure 2.3**
*Three subsystems with hypercyclic integration in a globalized dimension*

The network structure operates globally by constraining and enabling local substructures. However, the overall structure cannot be completely perceived from any of the positions in the network since there is no center of coordination. As this structure operates in a virtual dimension, it remains latent and cannot fully be observed locally. However, it can be hypothesized and then also measured. The theoretical specification of this virtual dimension reflects the evolving system.

## *2.2 Methods and materials*

As noted, the CD-Rom version of the *SCI 2000* contains 1,432,401 corporate addresses. These addresses point to 725,354 records contained in this database on a total of 778,446. Only 53,092 records (3.7%) contain no address information. Our current research focuses on the international coauthorship relations in this data, but we will report on that project elsewhere (Wagner & Leydesdorff, 2003). We focus on University-Industry-Government relations in this data set.

The addresses were organized in terms of their attribution to university-industry-government relations. This was done by a routine that first attributed a university label to addresses that contained the abbreviations "UNIV" or "COLL" in the address field. The remaining addresses were thereafter subsequently labeled as "industrial" if they contained one of the following identifiers "CORP", "INC", "LTD", "SA" or "AG". Thereafter, the file was scanned for the identifiers of public research institutions using "NATL", "NACL", "NAZL", "GOVT", "MINIST", "ACAD", "INST", "NIH", "HOSP", "HOP ", "EUROPEAN", "US", "CNRS", "CERN", "INRA", and "BUNDES" as identifiers. The order is so that hits are removed when retrieved using these routines. For example, addresses at the academy ("ACAD") cannot be confused with a university address, since the latter addresses have then already been removed.

This relatively simple procedure enabled us to identify 1,239,848, that is 86.6% of the total number of address records, in terms of their origin as "university," "industry," or "government." The distribution are as follows:

|  | *Number of addresses* | *Percentage* |
|---|---:|---:|
| "University" | 878,427 | 61.3 |
| "Industry" | 46,952 | 3.3 |
| "Government" | 314,469 | 22.0 |
| – (not identified) | 192,553 | 13.4 |
| *Total* | 1,432,401 | 100 |

**Table 2.1**
Addresses indicating university, industry or government affiliations in the *Science Citaiton Index 2000*

These sets can now be combined with country names. For example, of the 251,458 records containing an address in the U.S.A., 92.5 % (232,571) can be identified in terms of their origin in at least one of the three helices. More than 200,000 of these records (> 80%) contain at least one university address (Godin & Gingras, 2000).

|  | number | % ti | T(uig) in mbits | UI | UG | IG | UIG | Univers | Industry | Govern |
|---|---|---|---|---|---|---|---|---|---|---|
| all | 676511 | 93.3 | -77.0 | 16270 | 108919 | 4359 | 5201 | 543123 | 41242 | 232096 |
| USA | 232571 | 92.5 | -74.4 | 7200 | 37834 | 1782 | 2666 | 200149 | 18154 | 66416 |
| EU | 257376 | 93.0 | -50.1 | 4455 | 52112 | 1485 | 2028 | 206747 | 11192 | 101545 |
| JAPAN | 67715 | 97.9 | -92.1 | 4147 | 12492 | 954 | 1311 | 56534 | 9732 | 21664 |
| UK | 68404 | 93.1 | -63.1 | 1719 | 13098 | 394 | 690 | 54823 | 3970 | 26202 |
| GERMANY | 61017 | 94.7 | -43.4 | 1028 | 14003 | 407 | 664 | 51283 | 2799 | 23701 |
| FRANCE | 41112 | 90.3 | -52.1 | 439 | 11593 | 452 | 530 | 26133 | 1928 | 26595 |
| SCAND | 30939 | 95.8 | -31.6 | 490 | 8477 | 162 | 371 | 26542 | 1263 | 13005 |
| ITALY | 28958 | 89.9 | -29.4 | 362 | 7133 | 87 | 262 | 25633 | 905 | 10526 |
| NETHERL | 18357 | 95.3 | -25.4 | 372 | 4482 | 106 | 259 | 16379 | 863 | 6593 |
| RUSSIA | 22767 | 98.6 | -24.2 | 76 | 6315 | 162 | 138 | 11507 | 478 | 17611 |
| INDIA | 10916 | 89.2 | -78.1 | 97 | 1813 | 61 | 55 | 6099 | 407 | 6492 |
| BRAZIL | 9120 | 91.0 | -22.4 | 137 | 1727 | 32 | 52 | 7968 | 267 | 2885 |
| internat. coauthored | 120086 | 98.9 | -21.9 | 4550 | 47054 | 1349 | 2545 | 107569 | 9422 | 61138 |

**Table 2.2**
University-Industry-Government addresses and relations in the *Science Citation Index 2000*

What does this table teach us? First, it confirms that industry is not prominently present among the addresses of papers in the *Science Citation Index*. At the level of the database, industry is represented in appr. 6% of the papers. For the U.S.A. this figure is appr. 8%, and it is larger than 14% for Japan. However, this percentage is much lower for EU countries (4.3%). For example, this ratio is only 3.1% for Italy.

The table shows that in France the number of papers with addresses of public research institutes is larger than those with university addresses. This contributes to the triple-helix type of integration of the national system. $T_{U-I-G}$ is more negative for France than for Germany. The table shows that countries differ widely in terms of how the institutional arrangements operate among the main carriers of the knowledge infrastructure. The most negative value for the mutual information in three dimensions is found for Japan; the least negative for the internationally coauthored papers that are distinguished as a separate category in the last line of Table 2.2.

These results raise interesting questions that we will elaborate upon in another context. The main point here is that the data is useful in raising sophisticated questions like whether, and how, to measure and evaluate triple helix configurations. This measurement can also be combined with specific journal sets indicating disciplines and specialties. The mutual information provides us with an interesting indicator for the measurement of configurations within these sets and for the relations among them.

## *2.3   Webometric data*

Since the mid-nineties, a growing body of literature has emerged about measuring science and technology activities on the Web using informetric, bibliometric, and scientometric methods. In 1997 the name "webometrics" was introduced (Almind & Ingwersen, 1997). The journal *Cybermetrics* was also launched in 1997. Since then, the informetric community has taken up the investigation of the new electronic media, including the Internet (Larson, 1996; Rousseau, 1997; Ingwersen, 1998; Egghe, 2000; Thomas and Willet, 2000; Bar-Ilian, 2001; Bjöneborn and Ingwersen, 2001; Cronin, 2001; Thelwall, 2001).

If scholarly and scientific research and communication are more and more shaped by the Internet, analysis focussing on printed media may miss an important amount of research. In 1999, a first feasibility study granted by the European Commission stated "the opportunities for using informetric methods [on the Web] are not yet well elaborated" (Boudourides, Sigrist *et al.*, 1999). Meanwhile, several articles have appeared which tried to define the main topics of webometric approaches. Questions are raised such as: methods for adequate data collection and the use of search engines for that purpose (Snyder and Rosenbaum 1999); the problem of transferring terms like "citation" to the world of the Web ("sitations"; Rousseau, 1997); and the definition of impact factors for electronic journals (Ingwersen, 1998).

Let us provide an example of the possible use of Internet data by making a measurement effort comparable to the scientometric one outlined in the previous section. In a study of university-industry-government relations, Leydesdorff & Curran (2000) previously measured the occurrences and co-occurrences of the words "university," "industry", and "government" on the Internet using the AltaVista Advanced Search Engine. The advanced options of this search engine allow for the searching of various countries and general top-level domains (e.g., .com, .edu, etc.) in combination with specific time-frames for the publication dates of the websites, as well as Boolean operators.

Our previous study was replicated for different time periods, using various search engines by Bar-Ilan (2001). The author showed, among other things, how sensitive the Internet is for the measurement at different times (Rousseau, 1999). Here we will use the data only for the search terms "university", "industry", and "government", during the period 1993-2000. All measurements were performed on 13 November 2001, using the AltaVista Advanced Search Engine. The year 1993 was chosen as the first year of the time-series because web-based browsers based on hyperlinks were introduced at that time (Abbate, 1999).

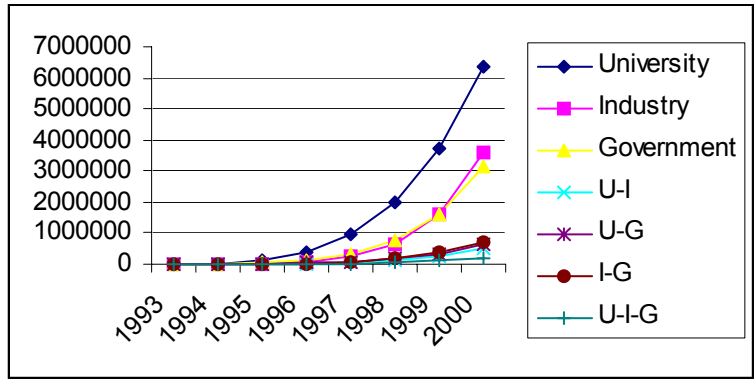

**Figure 2.4**
Results of searches using the *AltaVista Advanced Search Engine*

The data is organized in a three dimensional array, using the three search terms as independent dimensions, for each year as follows:

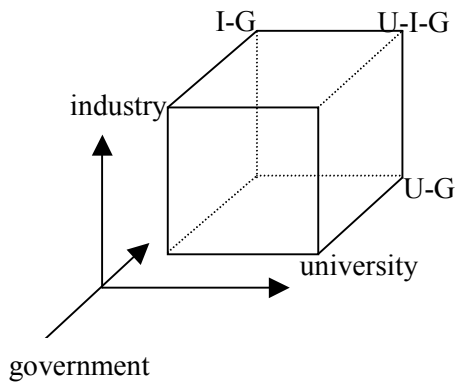

**Figure 2.5**
*A representation of university-industry-government relations
in a three dimensional array*

The values for T(uig) are always negative in the case of these Internet data, but the curve further decreases linearly since 1995 (Figure 2.6). It has previously been noted that the Internet experienced commercialization from 1995 (Abbate, 1999) and that the behavior of curves changed dramatically from that year onwards (Leydesdorff, 2000b). This and/or the rapid growth obviously leads to a further differentiation of the sets containing the three keywords as retrieved by the AltaVista search engine. The decrease is remarkably steady ($r^2 = 0.98$).

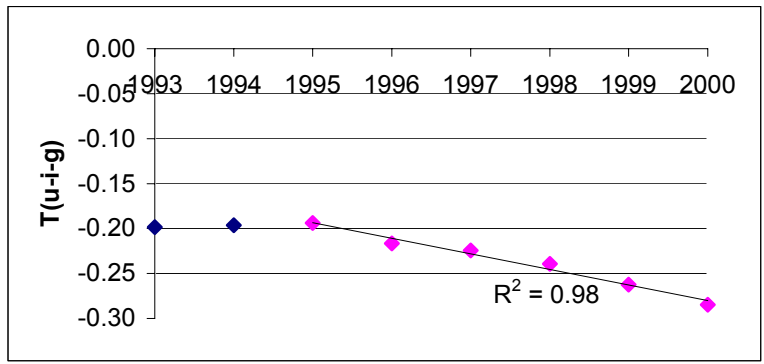

**Figure 2.6**
*Mutual information in three dimensions ("university," "industry," "government") as measured using the AltaVista Advanced Search Engine (at Nov. 13, 2001).*

## 2.4 Testing for Systemness

What does the above effect indicate in relation to the original data as exhibited in Figure 2.4 above? Does it really indicate the self-organization of a virtual dimension in the overlay of relations generated by the co-occurrences of two words in bi-lateral relations? Is this an indication of increasing self-organization of an overlay system?

For testing longitudinal data on whether the combined time series exhibit systemness in the data or not, a test was discussed in Chapter One. This test is based on evaluation of the Markov hypothesis for the collection of data versus the individual time series. The application of the test on this data provided the following results:

| prediction of the value in 2000 | 7 categories (U, I, G, UI, UG, IG, UIG) | four categories (UI, UG, IG, UIG) | three categories (UI, UG, IG) |
|---|---|---|---|
| on the basis of the univariate time series (1993-1999) | 1.18 | 9.81 | 10.67 |
| on the basis of the previous year (1999) (Markov property) | 8.84 | 1.87 | 1.65 |
| hypothesis of systemness | - 7.76 (rejected) | 7.44 | 9.02 |

**Table 2.3**
*Testing the hypothesis of systemness in the Triple Helix overlay of University-Industry-Government Relations. (All values are provided in millibits of information.)*

When reading this table, one should keep in mind that an observation does not generate any probabilistic entropy when compared with a perfect prediction. The value of the indicator, therefore, correlates negatively with the quality of the prediction. The results then show that the prediction of the 2000 data, on the basis of the same data for the previous year, is inferior to the prediction on the basis of the time series of the various categories in the case of considering the whole system of seven categories. Thus the hypothesis that these seven categories develop as a system is rejected.[6]

If the analysis is limited to the three bi-lateral relations (right column of Table 2.3), the hypothesis of systemness in this data is strongly corroborated. However, this prediction, is devalued by including the trilateral relations (middle column). So the conclusion is that this system of representations has been developing as a set of bilateral relations that contains a negative expected information value and, in this sense, self-organizes the complexity in the data using a virtual overlay of mutual relations.

## *2.5 Conclusion*

In the first part of this chapter, we showed how the mutual information of Triple Helix relations varies among world regions and countries. Data and statistics were provided at the level of the comprehensive database, but also more specifically for subsets indicating various countries (e.g., the USA, the UK, Russia, Japan) and regional blocks (e.g., the EU). The results raise interesting questions. For example, one can wonder why Germany deviates from other countries in its research portfolio as exhibited in Table 2.2.

Analogously, one can analyze international co-authorship relations as another boundary spanning mechanism, namely among nation states. Among other things, the subset of internationally co-authored papers was compared above with the datasets for individual countries. It could be shown that the internationally coauthored papers are far more homogenous in their pattern of cross-sectoral collaboration than any of the national sets (Wagner, 2002).

These results bring us to a third set of questions—still to be investigated at this moment in time—namely, the relationship between various boundary spanning mechanisms that can be analyzed within the data, for example:

1. international co-authorship data,
2. co-authorships across university-industry-government boundaries, and
3. co-publication in the same journal indicating an intellectual boundary spanning mechanism.

Note that boundary-spanning mechanisms operate by definition in a distributed and therefore uncertain mode.

---

[6] The best predictions from the individual time series are based on the last two years only, indicating in itself the rapid development of the Internet that tends to overwrite previous historical data as it develops. This is reflected in the exponential growth rates visible in Figure 2.4.

This further analysis would enable us to specify the relative importance of the sectoral differentiation in Triple Helix patterns, the intellectual organization in terms of journals, and the national subdivision of the system of international publications. Similarly, at the level of the Internet, one can repeat the above searches for specific domains like .nl for the Netherlands or .de for Germany.

These various dimensions can also be studied in terms of the mutual information among them. The results can then be interpreted, enabling us to raise further research questions. On the basis of previous projects we expect to find self-organization (that is, negative entropy in mutual information statistics) particularly when the dataset is restricted to addresses in the U.S.A., the European Union, and Japan, but not (or much less so) when we focus on relations among the 15 EU nations (Leydesdorff & Heimeriks, 2001).

As industry was poorly represented in the data from the *Science Citation Index*, the scientometric results were here above also compared with webometric data using the Altavista Advanced Search Engine. Our main purpose was to show the practicality of the methodology in both static and dynamic designs. On the Internet, a strong development of the Triple Helix could be demonstrated during the period 1995-2000. Using this specific representation, it could be shown that the development was self-organizing because of the bi-lateral relations between universities, industries, and governments. The unilateral and tri-lateral developments did not contribute to integration in this case.

The various data used in this study are interesting in themselves, but they remain statistical and, therefore, there are problems with the measurement (for example, when using AltaVista; Rousseau, 1999). We used the data sets above as examples of the sort of results that one is able to obtain when performing empirical Triple Helix or "Mode 2" research. Our argument in this study is that one can easily obtain complex data, but that these can only be analyzed with reference to a system that is hypothesized as being codified. In other words, the focus in this study was not on the measurement, but on the methodology, for analyzing data gathered for the evaluation of Triple Helix developments that have been measured—qualitatively and/or quantitatively—in terms of bi- and trilateral relations. A design for organizing this data and methods for its evaluation was specified.

**Chapter Three**

# THE COMMUNICATIVE TURN IN

# THE STUDY OF KNOWLEDGE-BASED SYSTEMS

In a complex and non-linear dynamic, each system remains under reconstruction and in evolutionary competition while reorganizing complexity within its relevant environments. Since communication systems are increasingly knowledge intensive, this reconstructive dynamics is continuously reinforced. The borders of the systems under study are then increasingly uncertain and therefore a subject of theoretical reflection. New codifications reconstructed by the ongoing processes of innovation and translation may become more functional than the underlying ones in an evolutionary mode. The previous configurations can be translated and partially overwritten.

The philosophy of science has been responsive to these developments in, and interfacing of, scientific communication. First, the systematic use of science in industry in the late 19$^{th}$ century raised fundamental questions about the *demarcation* between science and non-science at the interfaces. This issue led to the so-called "linguistic turn" in the philosophy of science during the interbellum. While truth had previously been associated with ideas, a truth-value was, henceforth, attributed to statements, with some being more likely to be true than others.

The "communication turn" has changed the situation once more. The truth-value of a statement can increasingly be considered as also contextual. One has a degree of freedom to play with the centrality of concepts in terms of heuristics and puzzle-solving (Simon, 1969 and 1973). Kuhn (1962), for example, noted that the precise definition of "atomic weight" differs between chemical physics and physical chemistry, without creating confusion. Concepts have meaning within discourses; meanings can be considered and reconstructed. Translations between discourses and reformulations can thus be considered as the carriers of knowledge-based developments (Leydesdorff, 2002).

This implies neither arbitrariness in what is true or not, nor a relativistic position. It implies an empirical orientation; communications leave traces that can be used as indicators. The various dimensions of a communication (including its potential truth value) can be distinguished. Values in these different dimensions can be measured when the communication system is fully specified as the hypothesis.

For example, scientific discourses can be expected to develop over time and thus may change in terms of what is considered to be true. Although the delineation of a discursive system (e.g., the paradigm) remains uncertain in terms of its boundaries, it can also be expected to be more certain in terms of its core. Codifications structure the discourses; translations enable the skilled participants to communicate among them. All these communicative acts can be observed and measured in terms of their intensity and frequency.

## 3.1 The need for "reflexive indicator research"

We argued that future indicator research should combine theoretical, historical, and empirical orientations. Indicators indicate communication processes that can (i) be analyzed substantively, (ii) modeled, and (iii) measured in a variety of dimensions. Hitherto, the focus in most analyses has remained on how communications develop historically. Starting at a certain point in time, paths of development can be traced following the arrow of time. The historical perspective dates the point for reference back in history.

The evolutionary perspective provokes the historical analysis by taking the present state as its point of reference. The system under study develops in the present by communicating given its past. One can then discuss how far one has to track back in order to understand the present state. The study of communication systems requires both historical and evolutionary oriented research designs because the evolutionary "incursion" (Dubois, 1998) takes place in history. Giddens (1979) has called this "a double hermeneutics." The understanding of the communication with hindsight feeds back on the historical understanding.

In addition to the historical and evolutionary analyses, the structural dimensions can be analyzed using sociological methods or the evolutionary metaphor of "variation and selection." This provides a snapshot of the complexity at one or another moment in time, e.g., the present. Note that the historical analysis tends to use concepts such as "change and stabilization" along the time axis, whereas structural analyses focus on "variation and selections" at specific moments in time. The same events ("variation") can be provided with different semantics by using different (orthogonal) axes for the reflection.

If one considers the three perspectives—the forward, the structural, and the backward analysis—as independent dimensions of reflection one generates a picture in which unstable phases and the emergence of new systemness appear much clearer than when using only a single (e.g., historical) metaphor for the representation. Such a reflexive (re)combination additionally provides the perspective of prospective intervention and policy-making because the recombination includes the discourse of what the representations may mean in the present.

In other words, one expects different perspectives on communication processes to be possible because a communication system—that operates in terms of changing distributions—can be accessed from different angles. The historical description of why specific patterns of variation and stabilization were shaped and reproduced, is one among these possible angles. The functions of communications, however, relate to the structures prevailing in the present. These can be analyzed sociologically, economically or sociometrically. However, the network representations may easily become too static.

They allow only for comparative statics. How can the underlying communication structures also be changed, for example, by policy and/or management interventions? How do options of change relate to structural patterns of ongoing processes of change (or drift)? An evolutionary perspective that appreciates the dynamics of stabilization with reference to globalization, can be added to the model.

A dynamic is complex insofar as it can be decomposed in terms of various interacting subdynamics. The variety of perspectives can be combined intuitively (for example, by a politician) and/or one can recombine the theoretical perspectives by using a model. The various perspectives have then to be formulated as manifestations of the complex system under specific conditions. One can try to capture this complexity by using a simulation model. However, the simulation model abstracts from the content in the underlying processes. It provides us with a formal representation that requires substantive appreciation by a reflexive discourse.

The social process is non-trivial: it makes each representation one knowledge claim among the possible representations. Latour (1988) has called this "infra-reflexivity." One cannot achieve a meta-position by further legitimating and/or delegitimating positions under study, for example, by using sophisticated mathematics. The formalization only enables us to evaluate the relative quality of the narratives as hypotheses explaining the phenomena and then the results may help to update the hypotheses for a next round of translations.

The unforeseen side effects, unintended consequences, etc., can be expected to challenge the various discourses to update given the yet unexplained and perhaps counter-intuitive findings of the quantitative evaluation. The research process continuously improves on the quality of the representation in a competitive mode, but the representations refer to systems that are developing at the same time. The knowledge-based system remains in transition and the study of the system therefore needs the further development of its reflections.

*The Organization of Reflexivity within the Systems*

The innovated systems absorb knowledge by being innovated. The observable arrangements therefore have an epistemological status beyond merely providing the analyst with one or another, as yet unreflexive starting point for the narrative. The data can be used for informing *ex ante*—and for sometimes testing *ex post*—the theoretical expectations. Which layer operated with which function, why, to which extent, and in which instances? This research program begins with expectations as different from observations: methodologically controlled observations can then inform the theoretical expectations.

The feedback layers of reflexive R&D management, science policy, citation analysis, etc. have in the meantime changed processes of knowledge production in science, technology, and society (Wouters, 1999). The nature of these changes is not unambiguous, but a number of hypotheses have been elaborated in the literature. However, there can be no doubt that new communication processes have been taking momentum since the introduction of the Internet and other information and communication technologies. Processes of communication can be expected to change basic mechanisms of knowledge production and communication. *The communication of knowledge feeds back on its production by further codifying and potentially changing previous knowledge.*

Indicator research so far has reacted sensitively on the changes in knowledge production. New indicators have been proposed regularly. The growing field of webometrics has witnessed an "indicator flood" in an increasingly information rich and knowledge-based environment. This creativity of indicator research may turn into a weakness if no theoretical backing can be developed. Which indicator is a relevant one, for which process, at which moment in time? Only few studies have tried to include a reflexive level when a new indicator is proposed.

What is indicated by the indicator and why is this indicator more suited to that purpose than comparable ones? There is an intrinsic need of validation studies within the indicator domain that is reflexive on the dynamics of the systems that are indicated. Reflexivity gains a particular urgency in phases of de-stabilization, re-organization and the emergence of new (and potentially innovative) structures. When the communication structures are developing at the same time, the starting points or the systems of reference have to be made as clear as possible so that one can trace the changes that are under study in relation to the changes that are made visible and/or explained by the study. This reflexivity can be elaborated in each of the three dimensions: theoretically, historically, and empirically.

By raising first the substantive question of "what is communicated?"—e.g., economic expectations (in terms of profit and growth), theoretical expectations or perhaps scenarios of what can technologically be realized given institutional and geographic constraints—the focus is firmly set on the specification of the media of communication. How are these communications related and converted into one another? Why are these processes sometimes mutually attractive and reinforcing one another, and under what conditions can the exchanges among them be sustained?

In the historical dimension reflexivity stands for the introduction of a perspective that focuses on dynamic processes in contexts rather than on historical results (e.g., Barnes & Edge, 1982; Latour, 1987). As noted, the evolutionary perspective includes the time axis, but as a degree of freedom. The present is the relevant system of reference for policy analysis. However, the present is also historical, that is, as a transient state towards new developments. Furthermore, the reflexive analyst is aware of one's own position in relation to previous lines of research and one's social contexts.

In the empirical dimension, reflexive indicator research also communicates the expected boundaries of the methods and data used. Which were the selection criteria? What would count as unexpected events? Can the surprise value of the newly emerging developments be expressed in bits of information added to the system? Thus characterized "reflexive indicator research" is not a new paradigm. It reflects traditional standards of scientific analysis. But, given the drive by the data in indicator research and the development towards an algorithmic understanding, the strengthening of a theoretical approach may become a necessary condition for the further development of quantitative approaches to the study of science, technology, and innovation.

During the last two decades, the qualitative and the quantitative traditions in science and technology studies have grown increasingly apart (Leydesdorff & Besselaar, 1997; Van den Besselaar, 2000 and 2001). It is time for the pendulum to be turned given the urgent need to understand the effects of different forms of communication and their interaction in knowledge production. In our opinion, the growing diversification and specialization in the sciences, and the relationships to their societal environments, in a knowledge-based economy calls for integrative approaches with detailed appreciation of the ongoing processes of differentiation. It is only on such a basis that one can more precisely describe the options for making choices both in the public and in the private domains (of enterprises, research groups, etc.).

The need for knowledge-based science-policy making to be able to make a distinction between what might make a difference and what might not, is reflected in the seriousness of the problem of integration and differentiation in the theoretical description and explanation of the knowledge-based systems under study. Both qualitative theorizing and quantitative information are then needed. Our theoretical framework is neither exclusive nor normative. On the one hand, we need the qualitative contributions because they generate hypotheses. On the other, indicator researchers can pay attention to the elaboration of the theoretical frameworks implied in their research. As communication theoretical, systems theoretical, and evolution theoretical concepts are involved, this task of integration through reformulation cannot be considered as a *sine cura* (Luhmann, 1975).

It is a widely held prejudice that quantitative analysis is data-driven and poor in theorizing. In our opinion, this is a cultural misunderstanding. No measurement is purely technical; theoretical baselines are always involved. Each definition of a variable implies an image of a process that is represented. The theoretical references are often not completely described in quantitative studies. However, the problem is not "missed theory," but "invisible theory."

The cure is discursive reflexivity. Indicator research is not a discipline with a single and commonly accepted theoretical background. It is an "interdiscipline" with approaches as different as the disciplinary background of the researchers.[7] What can still be justified in the case of research results presented to one scientific community of specialists—when the theoretical foundations provide a common basis so that they do not have to be repeated—may loose its justification when crossing a (sub)disciplinary boundary.

If the theoretical backgrounds in indicator research are not sufficiently reflected, it becomes impossible to create "trading zones" (Nowotny et al., 2001). These trading zones are needed in order to create a dialogue between different approaches inside the branch of quantitative analysis (e.g., between simulation studies and measurement efforts) as well as towards qualitatively oriented science, technology, and innovation studies. New research questions can then be formulated that appreciate the previously achieved results.

---

[7] The observable data can be considered as "phenotypical," whereas the perspectives for their interpretation compete as "genotypes" that may be able to explain the observable variations (Langton, Taylor, Farmer, & Rasmussen, 1992).

## *3.2  Indicators as representations of codified communications*

Our theoretical contribution in this study has been the use and operationalization of the communication-theoretical framework for studying reflexively developing systems of knowledge production and control. This perspective enabled us to understand an indicator as a specific representation of a process of knowledge production and communication. The process that is represented can be specified as a theoretical hypothesis, and the indicator then provides us with observations that can be used to inform (enrich or sometimes reject) the hypothesis.

In general, a STI indicator stands for a social process in science and technology. The processes of communication can be made observable by using the indicator. However, since the processes of communication are distributed, the measurement can be expected to contain an uncertainty. The observations therefore have to be interpreted.

We use a communication theoretical and systems theoretical approach. Social processes become visible in the communications used in the social systems under study. Following Luhmann (1984) and others, social communication systems can additionally be expected to differentiate functionally. Accordingly, the communications develop different systems of communications endogenously. We used Luhmann's notion of "codification" to describe the different forms in which communications can be expressed, stored, and recalled. Indicator research is based on the assumption that it is possible to recall the information more precisely by methodologically controlling the measurement instruments.

The "literature model" has dominated the quantitative study of scientific communication in scientometrics for many decades. The scientific journal article has been considered as the core of this model and, as a result, codified communication was the basic form of communication under study. However, the changes in knowledge production and its embedding in communication, as we observe these phenomena nowadays, have consequences for the codification of scientific communication.

For example, the increased use of information and communication technologies in science (e.g., on-line publications, digital data production, and simulations) may already have challenged the model of the "journal article" as the prevailing form of scientific communication. The shift of attention from science, to science-based technology and innovation has led to systematic indicator research in patent databases. The embeddedness of science in technology, and vice versa, can be traced analyzing the differences and asymmetries in citation patterns between the domains of scientific literature and patenting (Schmoch, 1997; Grupp & Schmoch, 1999; Meyer, 2000a,b).

In order to produce a measurable indicator, the hypothesis of a process has to be operationalized. Variables are then defined. Measurement may lead to repeatable and reliable results (or not). The availability of databases functions as a constraint. Databases are the backbone of indicator research. Both the availability of databases and the development of statistics provide inherent limitations to indicator research.

In this paper we mainly focused on databases like the *Science Citation Index*. *Medline* and the *European Patent Office* database are other examples of data bases used in bibliometrics. We also discussed web data, e.g. using the Advanced Search Engine of *AltaVista*. In general, dedicated databases can be understood as representing specific types of codification in knowledge-based communication. For example, the *Science Citation Index* is mainly useful for mapping the communication in science located in universities and public research institutions. Patent databases represent communications about technologies. The classification scheme outlined in Table 3.1 may be helpful in organizing the respective domains. Our emphasis on differentiation at interfaces as a condition for innovation has led us to focus on the combination of the different data sources.

|  | University | Government | Industry |
|---|---|---|---|
| Science | | Science Citation Index | |
| Technology | | Patent data bases | |
| Innovation | | | Market data |

**Table 3.1**
*Functional versus institutional differentiation in the Internet age*

We consider the process of knowledge creation as a stepwise process from so-called "fundamental" knowledge towards market relevant innovations, and vice versa, from the market into the knowledge production process. This process contains feedback loops within each of the systems and among them. Each subsystem develops recursively and interactively. The feedback loops control the forward movement of the process in an organized way (Kline & Rosenberg, 1986).

The interaction of different knowledge networks links different phases in a heterogeneous process. Each heterogeneous process itself contains one reconstruction of the historical events among others possible, but the selection takes place from a hindsight perspective. It can be considered as an actualization (a state) of the system. The different phases and processes can be made visible as differences in the codifications. The focus remains on the emerging systems that result from this non-linear dynamics.

A linear combination of databases only increases the complexity of the description by extending a relatively simple representation into the multidimensional perspective of interacting subdynamics. In order to handle the complexity, theory has to be introduced as an integrative and organizing element. The emerging system can be hypothesized.

For example, one can consider the case of the creation and introduction of a pharmaceutical to exemplify how multidimensionality can be bundled together in the description of the specific process of knowledge creation and innovation (Leydesdorff, 2001b). The innovation can be analyzed as a performative act in history (Latour, 1987). Our perspective, however, is (neo-)evolutionary and systems theoretical: how are the coordination mechanisms between functional domains affected by innovation? The knowledge-based innovations can be expected to reconfigure the structures on which they build by reconstructing and recombining them in terms of new representations.

The uncertain definition of a system of innovation in terms of nations, sectors, technologies, regions, etc., brings players other than the traditional ones into scope. Following upon the Bayh-Dole Act (1980), for example, universities in the U.S.A. have been stimulated to submit patent applications. Does university research already play a strategic role in a domain of patenting? Whereas this role can historically be analyzed for innovations on a case-by-case basis, the delineation of a system of innovations is required for defining this role at the aggregated level.

*We propose to consider the study of "innovations" and the potentially systemic character of clusters of innovations as **a third program of research** in science, technology, and innovation studies.* Science indicators have hitherto focused on performance and scientific impact. Patent indicators measure technical inventiveness from a historical perspective (Sahal, 1981). Innovation indicators turn the tables by using a hindsight, systems, and/or evolutionary perspective. Innovation is per definition an emerging unit of analysis based on communication between different systems. The innovated systems can be changed significantly by an innovation.

Innovations have been analyzed mainly under the aspect of technological diffusion and technological forecasting. Note that the system of reference of such studies has been the technological development under study: One then asks for the consequences or impacts of new technologies, for example, in terms of technology assessment. In innovation studies, technological developments (or stagnations) themselves have to be explained. Under which conditions can further innovation be expected? Unlike bibliometric and patent analysis, modeling plays a more dominant role in this area because of the focus on new and emerging options.

Empirical studies that trace the growth of an innovation back have been relatively scarce in science and technology studies and evolutionary economics (Von Hippel, 1988). One reason may have been the lack of standardized "innovation databases" (Pavitt, 1984). Data gathering is often time-consuming, being tailor-made (Frenken, 2000 and 2001; Frenken & Leydesdorff, 2001). The Internet provides a new perspective on "data mining." Systematic links between innovation (market), invention (patent) and scientific knowledge (literature) can now be constructed.

A closer connection between science, technology, and innovation studies has also theoretical consequences. In innovation studies the focus is on the (re)constructed system as different from the historical construction. Models and simulations are introduced to explore the dynamical nature of the reconstruction. From the perspective of the reconstructed system, innovation potentially restructures the history of the representations because the system continuously selects upon the variety of possible representations for its reconstruction.

For example, when a national system of innovation is assumed as the system of reference, dimensions can be appreciated other than when one assumes a sector (e.g., chemistry) or a new technology (e.g., biotechnology) as the evolving system. One can always question these delineations as assumptions; they can only be used as starting points for the reconstruction. Therefore, the definitions and delineations have to be communicated together with the indicators proposed. Different perspectives can be expected, since innovations take place at the interfaces between systems. *Reflexive* indicator research becomes necessary when innovations are made the focus of research.

On the basis of understanding the processes of knowledge creation as self-organizing and complex, we indicated above how one can test for the hypothesized phenomena in terms of new structures potentially reproduced by coherent behavior. The case of the emergence of a so-called European Research Area is such an example one would look for. The politically motivated proposal aims at an institutional innovation that should drive the European sciences into a phase transition. The emergence of new forms of self-organization can sometimes be tested (e.g., Leydesdorff, 2000a; Leydesdorff & Heimeriks, 2001).

This approach, that is of testing a hypothesis as a research question, can be extended to other theses. For example, the "Mode 2" thesis posits a change in the system of innovations in cognitive, social as well as institutional dimensions. To which extent can hypotheses based on the "Mode 2" theory be tested empirically by using indicator research? Independently of the acceptance of the model as an explanation, one can entertain the thesis as a hypothesis and ask for new social forms of knowledge production like virtual communities or collaboratories. What have been the effects of electronically mediated communications on scientific knowledge production and diffusion? In this context, web indicators are perhaps the most appropriate representations (Zelman, 2002).

*Web Indicators*

In the empirical part of this study, we drew attention to the possibilities and limitations of web indicators. The Internet represents a medium for differently codified communications. Traditional databases can nowadays be used on-line like the *Web of Science*—the on-line version of the *Science Citation Index*. Traditional communication channels like the scientific journals become increasingly available on-line.

In addition to these trends of digitalization, the web offers the possibility to trace research activities that have not yet been documented. The rise of so-called *Virtual Ethnography* (Hine, 2000) is only one among a variety of new methodologies that have become available in science and technology studies. Cybermetrics or Webometrics stand for quantitative approaches in this direction. Problems of reliable data and stability of the measurement over time are major methodological problems when using web indicators. These problems do not have to surprise us given the dynamic character of the Internet.

Using search engine or meta-crawlers, one can compare the frequency of very different kind of communications (keywords) as well as of institutions (e.g., host extensions). However, further methodological research concerning the stability of search engines may then increasingly be necessary. Different update frequencies can be expected in different domains. In summary, one can state that automatized ways of data mining have to be developed to use the "data flood" on the web for indicator purposes. The automation and consequent black-boxing of theoretical assumptions into standards, however, generates another tension that can drive new research processes both empirically and theoretically. The standards may have to be updated regularly.

The standardization of "purchasing power parity" by the OECD can be considered as an early example of an evolutionary indicator because the values of these input indicators had regularly to be updated with reference to changes in the exchange rates. This study concentrated on output indicators. However, we wish to draw attention to the effect of these further developments in output measurement on the study of input indicators. Input indicators like R&D expenditure, R&D personnel have mainly been developed by the OECD (e.g., 1976) and are often used in macro analyses of science policy (e.g., OECD, 1980).

Relatively less attention was drawn in this report to the effect of the ICT revolutions and the emerging focus on innovation on research of input indicators (OECD/Eurostat, 1997). For future indicator research, however, one may wish to raise questions about the need to match new (e.g., web-based) indicators with input indicators. For example, ICT is often not classified as R&D. How is the efficiency of spending in new areas of technoscience to be measured if interaction effects between "R&D" and "non-R&D" activities may become more important than the sum of the two efforts (Kaghan & Barnett, 1997)?

## *3.3 A Program of Innovations Studies*

A fundamental reformulation of the problems of Science, Technology, and Innovation Policies became urgent during the 1990s. Three models have been central to the discussion about studying innovation systems: (i) the proposal to distinguish a "Mode 2" type of knowledge production, (ii) the model of "national systems of innovation", and (iii) the triple helix of university-industry-government relations.

The authors of the "Mode 2" thesis (Gibbons *et al.*, 1994) have argued that the new configuration has led to a dedifferentiation of the relations between science, technology, and society. From the perspective of these authors, all scientific and technical communication can be equated and compared with other communication from the perspective of science, technology, and innovation policies.

In our opinion, this model is based on a confusion of the representation with the represented system under study. The political or managerial representation provides us with a window that can be integrated because it uses a specific medium of communication. However, the represented systems are operationally expected to remain differentiated. If the integration is also successful in the systems under study (e.g., in the case of a reconstruction or innovation), the systems are integrated at their interfaces and therefore they can be expected to restore also their own orders by differentiating again after the integration. The integration means something different for differently codified systems.

Differentiation and integration do not exclude one another, but rather assume one another. The communication enables us to construct an integrated picture, but the underlying systems compete both in terms of their social realities and in terms of the representations that they enable us to construct at the interfaces. Systems of innovations solve the puzzle of how to interface different functions in the communication at the level of organization.

Evolutionary economists have argued in favor of studying "national systems of innovation" as hitherto the most relevant level of integration. Indeed, they have provided strong arguments for this choice (Lundvall, 1992; Nelson, 1993; Skolnikoff, 1993). However, these systems are continuously restructured under the drive of global differentiation of the expectations. Economies are interwoven both at the level of the markets and in terms of multinational corporations, sciences are organized internationally, and governance is no longer limited within national boundaries. The most interesting innovations can be expected to involve boundary-spanning mechanisms.

In other words, we agree with the "Mode 2"-model in assuming a focus on communication as the driver of systems of knowledge production and control. However, the problem of structural differences among the communications and the organization of interfaces remain crucial to the understanding of a global and knowledge-based economy. The wealth from knowledge and options for further developments have to be retained by reorganizing institutional arrangements with reference to the global horizons.

The Triple Helix model of university-industry-government relations tries to capture both dynamics by introducing the notion of an overlay that feeds back on the institutional arrangements. Each of the helices develops internally, but they also interact in terms of exchanges of both goods and services and in terms of knowledge-based expectations. The various dynamics have first to be distinguished and operationalized, and then sometimes they can also be measured. Throughout this report we have tried to show how the dynamics between the dimensions can then be reconstructed using indicator research.

The strength of this research program is that it does not simply generalize on the basis of intuitions. The empirical results can be expected to inform us. As the complexity increases, the results may often be counterintuitive. One may be able to appreciate them by innovating one's theoretical assumptions. As the various subdynamics can better be understood, one may also be able to develop simulation models on the basis of their reconstructions.

There is an intimate connection between indicator research and parameter estimation in simulation studies when analyzing knowledge-based systems. Indicators study knowledge production and communication in terms of the traces that communications leave behind, while simulations try to capture the operations and their interactions. The common assumption of indicator research and simulation studies is that knowledge production, communication, and control are considered as operations that change the materials on which they operate. The unit of analysis is replaced with a unit of operation.

The difficult relations between empirical studies and algorithmic simulations have to be guided by theorizing. Otherwise, the number of options explodes without quality control. What do the different pictures mean? Both theoretical specification and methodological control are needed. In our opinion, the study of communication and the interfacing can use available theories of communication. We have argued that two theories are then particularly important: (i) Luhmann's sociological theory of communication with its emphasis on functional differentiation (Luhmann, 1984 and 1990; Leydesdorff, 2001a) and (ii) the mathematical theory of communication that can be used for the operationalization (Shannon, 1948; Abramson, 1963; Theil, 1972; Leydesdorff, 1995). The combination of these two theories with a very different status—that is as theory and methods—enabled us in the various chapters of this study to update and inform empirical hypotheses about how the knowledge base transforms the institutional relations of an increasingly knowledge-based society.

## 3.4  Policies of Innovation: Innovation of Policies?

The gradual transition from a political economy to a knowledge-based economy potentially changes the cause-effect relationships between the control systems and the systems to be steered. As the system to be steered becomes increasingly self-organizing—for example, in terms of containing "lock-ins"—the options for steering become dependent on the windows that the systems leave for intervention. These windows can only be established on the basis of knowledge-based reconstructions.

For example, in a political economy, the political system is inclined to steer a system like the scientific enterprise (or the national system of innovations) in terms of its institutional parameters (Spiegel-Rösing, 1973; Van den Daele, Krohn, & Weingart, 1977). In a knowledge-based economy, institutional parameters tend to lose their relevance as the institutions are under pressure of reorganization. Networks of institutions shape university-industry-government relationships in a non-linear dynamic. The latter are driven by political and economic opportunities to grasp competitive advantage.

The knowledge-base of the economy is highly structured by relevant interfaces that are continuously reproduced and yet differentiated within the system and its relations to different environments. "Validation boundaries" can, for example, be considered as a knowledge-based equivalent of institutional boundaries (Fujigaki, 1998). Validation boundaries are the result of codification processes that reconstitute institutional delineations. A validation boundary can be expected to have an internal and external side with different characteristics. Fujigaki & Leydesdorff (2000) have elaborated upon the concept of "validation boundaries" for exchange and control processes at the societal interfaces of knowledge-based systems.

Validation boundaries can also be considered as condensations and stabilizations of interacting communication processes. Each of the communication processes selects asymmetrically and asynchronically at relevant interfaces, but some selections can be selected for stabilization; some stabilizations can recursively be selected for globalization. The institutional level provides the stability that is needed for participation in the globalization. Thus the perspective of the institutional optimalization refers to the carrying capacity and the sustainability of the network arrangements.

In addition, the institutional environment provides a trade-off between the mechanisms of a political economy such as public control and private appropriation. If one begins the political process only at the latter end—for example, because that is the traditional routine of producing policy decisions—one tends to lose precisely the knowledge-based dimension in the policy-making and/or the managerial processes. Knowledge-based development requires the policy and management control process itself to be innovated accordingly.

For example, institutional interests have been shaped by history. We have argued that evolutionary analysis requires the consideration of the historical formations reflexively. The functionality of the institutions can continuously be discussed and analyzed. The discussion of the functionality of the delineations can be informed by indicator research.